\documentclass[longbibliography,noeprint,noshowpacs,nopreprintnumbers,twocolumn,pra,showpacs,superscriptaddress,amsmath,amssymb,citeautoscript,aps,10pt]{revtex4-2}
\usepackage{graphicx}
\usepackage{dcolumn}
\usepackage{color}
\usepackage{cancel}
\usepackage{bm}
\usepackage[hidelinks]{hyperref}
\hypersetup{
    colorlinks,
    citecolor=blue,
    filecolor=blue,
    linkcolor=blue,
    urlcolor=blue
}
\graphicspath{{Figures/}{./Figures/}{.}}

\DeclareMathOperator{\tr}{tr}
\DeclareMathOperator{\C}{\mathcal{C}}
\DeclareMathOperator{\hc}{\mathcal{H}}
\DeclareMathOperator{\hh}{\hat{H}}
\renewcommand{\vec}{\bm}

\begin{document}

\title{The Two Dimensional Dynamical Bulk Boundary Correspondence: Beyond Two Band Models}

\author{Tomasz Mas{\l}owski}
\affiliation{The Faculty of Mathematics and Applied Physics, Rzesz\'ow University of Technology, al.~Powsta\'nc\'ow Warszawy 6, 35-959 Rzesz\'ow, Poland}
\author{Nicholas Sedlmayr}
\email[e-mail:]{sedlmayr@umcs.pl}
\affiliation{Institute of Physics, M. Curie-Sk\l{}odowska University, 20-031 Lublin, Poland}

\date{\today}

\begin{abstract}
A dynamical equivalent of the bulk-boundary correspondence has been suggested to occur in one and two dimensional topological models following sudden quenches. Depending on the topological invariant of the time evolving and initial phases involved large boundary contributions to a dynamical free energy occur. Moreover they occur periodically between the critical times at which this dynamical free energy becomes non-analytic, \emph{i.e.}~at dynamical quantum phase transitions. At these critical times the eigenvalue spectrum of the Loschmidt matrix which underlies the dynamical free energy closes its gap. The boundary contributions are understood to be due to zero-modes or in-gap bands of this matrix, forming a close analogy with equilibrium topological models and their edge modes. The exact cause of this phenomena and its generality remain unknown. In this article we test the dynamical bulk-boundary correspondence for a complicated two dimensional topological superconductor with a rich phase diagram, allowing quenches between many different Chern numbers. We show that there is no straightforward correspondence between the equilibrium phases quenched between and the dynamical bulk boundary correspondence. Furthermore the correspondence can depend on the orientation of the edges, suggesting a possible weak topological variant.
\end{abstract}

\maketitle

\section{Introduction}\label{sec:intro}

The bulk boundary correspondence maintains a crucial position in the field of topological insulators and superconductors. It relates the topological classification of bulk bandstructures via topological indices~\cite{Ryu2010,Hasan2010,Trifunovic2021}, to the existence of states of lower dimensions at boundaries and defects~\cite{Chiu2016,Trifunovic2018}. These states have a robustness due to the bulk topological features and can not be easily destroyed without breaking the symmetries underlying the bulk topology. The topological index, and hence topological phase, can typically only be changed by closing the gap~\cite{Maska2021}. The topology itself is a property of the bandstructure, and is therefore useful in characterizing equilibrium properties of systems. What happens when the system is perturbed from equilibrium is less well understood and few general statements can be made. One way of studying non-equilibrium phenomena is to consider a sudden quench in parameters. This essentially results in considering initial states, defined typically as eigenstates of one Hamiltonian, which are time evolved by different Hamiltonians. Quenches are widely used for example in quantum thermalization studies~\cite{Rigol2008a,Polkovnikov2011,Rigol2012,Sirker2014}.

Following such quenches one possibility is for dynamical quantum phase transitions (DQPTs) to occur~\cite{Heyl2013,Andraschko2014,Heyl2018a,Heyl2019,Sedlmayr2019a}. DQPTs themselves are defined as non-analyticities in a dynamical free energy which occur at critical \emph{times}. They are therefore not necessarily associated with a precise notion of a dynamical phase. In the literature we note that the phrases dynamical phase transition and dynamical quantum phase transition have received a variety of definitions~\cite{Schutzhold2006,Eckstein2009,Diehl2010,Mitra2012a,Marino2022,Corps2022}. Here we focus purely on the concept of non-analyticities in the dynamical free energy, which already allow for physics not predicted by the equilibrium phases of the initial and time evolving Hamiltonians~\cite{Vajna2014,Andraschko2014,Karrasch2017,Jafari2017,Cheraghi2018,Jafari2019,Wrzesniewski2022} Connections of the DQPTs to other concepts of dynamical order~\cite{Heyl2013,Karrasch2013,Heyl2014,Budich2016,Halimeh2018,Hagymasi2019,Uhrich2020,Wrzesniewski2022,Bhattacharyya2026a} or to entanglement~\cite{Torlai2014,Sedlmayr2018,Gong2018a,Maslowski2020,Poyhonen2021,Bhattacharyya2026a} have been studied, but no definitive relation has been found.

In analogy to a partition function and free energy we can introduce the dynamical notions of the Loschmidt amplitude, the overlap between a time evolved and initial state, and the return rate, the scaled log of the Loschmidt amplitude. The Loschmidt overlap suffers an orthogonality catastrophe in the thermodynamic limit, hence the requirement for the return rate. The non-analyticities themselves, our DQPTs, occur at critical times in the return rate, visible as a cusp, and are caused by zeroes of the Loschmidt amplitude. The zeroes can be studied by the behaviour of Fisher zeroes. Generalizing the Loschmidt amplitude to complex times Fisher zeroes are the complex times at which the Loschmidt amplitude vanishes. For a one dimensional model they form lines in the complex plane which cause DQPTs when they cross the real time axis~\cite{Heyl2013}. In higher dimensions Fisher zeroes form two dimensional areas in the complex plane, and the occurrence of DQPTs depends on the behaviour of the density of zeroes crossing the real time axis~\cite{Maslowski2024b,Schmitt2015,Vajna2015}. In such a case the non-analyticity results in a cusp in the first derivative of the return rate at the critical time.

DQPTs have now been widely studied theoretically~\cite{Karrasch2013,Heyl2014,Heyl2015,Sharma2015,Halimeh2017,Homrighausen2017,Halimeh2018,Shpielberg2018,Zunkovic2018,Srivastav2019,Huang2019,Gurarie2019,Abdi2019,Puebla2020,Link2020,Sun2020,Rylands2021,Trapin2021,Yu2021,Halimeh2021,Halimeh2021a,DeNicola2021,Cheraghi2021,Cao2021,Bandyopadhyay2021,Cheraghi2023,Wong2023b,Zhang2026a}, experimentally~\cite{Jurcevic2017,Flaschner2018,Zhang2017b,Guo2019,Smale2019,Nie2020,Tian2020,Pomarico2026}, and the idea has been generalized beyond using initial pure states to finite temperatures and open systems~\cite{Mera2017,Sedlmayr2018b,Bhattacharya2017a,Heyl2017,Abeling2016,Lang2018,Lang2018a,Kyaw2020,Starchl2022,Naji2022,Kawabata2023}. The occurrence of DQPTs in topological models is also well established~\cite{Schmitt2015,Vajna2015,Jafari2016,Bhattacharya2017,Jafari2017a,Sedlmayr2018,Jafari2018,Zache2019,Maslowski2020,Okugawa2021,Rossi2022,Maslowski2023,Maslowski2024} and possible dynamical order parameters have also been proposed~\cite{Budich2016,Heyl2017,Bhattacharya2017a}. These works mostly focus on simple two band one dimensional models. Higher dimensional models have received less attention~\cite{Schmitt2015,Vajna2015,Bhattacharya2017,DeNicola2022,Hashizume2022,Brange2022,Sacramento2024,Kosior2024,Maslowski2024b}, as have many-band models~\cite{Huang2016,Jafari2019,Mendl2019,Maslowski2020,Maslowski2023,Maslowski2024,Maslowski2024b}.

Another limitation is the focus purely on bulk properties. As one of the principle motivations for studying topological models is the bulk boundary correspondence it is interesting to consider the role that the edge state can play in the dynamics. DQPTs in topological models have been shown to possess a dynamical bulk-boundary correspondence (DBBC) in one dimension~\cite{Sedlmayr2018,Sedlmayr2019a,Maslowski2020}, two dimensional higher order materials~\cite{Maslowski2023,Maslowski2024}, and simple two dimensional topological models~\cite{Maslowski2026a}. The DBBC can be understood in two ways. Firstly as large contributions to a suitable defined boundary return rate, which occur between successive critical times. These contributions only exist for specific cases depending on the topology of the time evolving Hamiltonian and initial state.

A more direct analogy to the equilibrium bulk-boundary correspondence can be drawn by considering the cause of the boundary return rate. The Loschmidt amplitude can be defined as the determinant of a non-Hermitian Loschmidt matrix, and therefore as the product of its eigenvalues. Critical times are zeroes, \emph{i.e.}~gap closings, of this spectrum of eigenvalues. The large boundary return rates, in one dimensional models and two dimensional higher order models, are caused by zero valued eigenvalues which occur between successive gap closings of the Loschmidt spectrum. The exact nature of the relation between the Loschmidt matrix and the zeroes is currently not known, but appears to be analogous to the zero modes of a topologically non-trivial Hamiltonian matrix. We note that the forms of dynamical order parameter so far defined~\cite{Budich2016,Heyl2017,Bhattacharya2017a}, do not predict the DBBC. For the two dimensional case the natural extension, \emph{i.e.}~dispersive in-gap bands, occur between critical times and cause distinctive patterns in the boundary return rate~\cite{Maslowski2026a}.

In this article we extend the idea of the two dimensional DBBC beyond the simplest two band modes with limited equilibrium phase diagrams. By considering a more complicated model of a topological superconductor on a honeycomb lattice~\cite{Sedlmayr2017} we allow for two possible generalisations. The first is the availability of a wider range of Chern numbers for the equilibrium phases to quench between. The second is to have (at least) two distinct types of boundary easily accessible for which the boundary effects can be compared. By considering all these possibilities we find that although the DBBC survives, there is no simple relation to the equilibrium tropological phase diagram. We also see hints of a form of weak dynamical topology. Due to the complications of this model it is not possible to extract the boundary return rate from a finite size scaling analysis, and hence we study the Loschmidt matrix spectrum directly.

This paper is organized as follows. In section \ref{sec:dqpt} we introduce DQPTs and the DBBC. In section \ref{sec:mods} we define the topological superconductor model we will study as an example. In section \ref{sec:hexdbb} we show the results for the dynamical bulk-boundary effect and discuss its general properties. In section \ref{sec:con} we conclude.

\section{Dynamical Quantum Phase Transitions and the Dynamical Bulk-Boundary Correspondence}\label{sec:dqpt}

We focus here on quenches for which we have an initial state $|\Psi_0\rangle$, which is the ground state of a Hamiltonian: $\hh_0|\Psi_0\rangle|=E_0|\Psi_0\rangle$. This state is time evolved by a second Hamiltonian $\hh_1$ and the Loschmidt amplitude itself is defined as
\begin{equation}
L(t)=\langle\Psi_0|e^{-i\hh_1t}|\Psi_0\rangle\,,
\end{equation}
for a system of $N$ unit cells. The related return rate is
\begin{equation}\label{return}
l(t)=-\frac{1}{N}\ln|L(t)|\,,
\end{equation}
which remains well defined in the thermodynamic limit despite the orthogonality catastrophe which plagues the Loschmidt amplitude. In the thermodynamic limit we define $l_0(t)\equiv\lim_{N\to\infty}l(t)$. We have a DQPT when $l(t)$ becomes non-analytic as a function of time. To understand when this occurs we consider $L(-iz)$ for complex $z$ and study the behaviour of the Fisher zeroes $L(-iz^*)=0$ in the thermodynamic limit. For a two dimensional model Fisher zeroes form areas in the complex plane. A DQPT occurs when, along the real time axis, there is a discontinuity or divergence in the Fisher zero\textcolor{red}{es} density~\cite{Schmitt2015,Vajna2015,Maslowski2024b}, and the result is a cusp in the derivative of the return rate:
\begin{equation}\label{returnd}
\dot l(t)=\frac{\partial l}{\partial t}(t)\,.
\end{equation}
For the model we will consider in this article direct calculation of the Fisher zeroes is not feasible, and we also relegate discussion of the bulk DQPTs to appendix~\ref{app:bulkdqpts}.

As we are interested in boundary effects we need to calculate the Loschmidt amplitude for open systems. For this purpose we use a method based on the correlation matrix, $\C_{ij}=\langle\Psi_0|\Psi^\dagger_i\Psi_j|\Psi_0\rangle$, where $i,j$ label a full basis of creation and annihilation operators. Now using~\cite{Levitov1996,Klich2003,Rossini2007,Sedlmayr2018}
\begin{equation}\label{rle}
    L(t)=\det{\bm M}(t)\equiv\det\left[1-\mathbf{\C}+\mathbf{\C}e^{it {\mathcal H}_1}\right]\,,
\end{equation}
one can calculate the Loschmidt amplitude or properties of the Loschmidt matrix $\bm M(t)$, where ${\mathcal H}_1$ is the Hamiltonian density. The eigenvalues of $\bm M(t)$ are $\lambda_i(t)$, in terms of which
\begin{equation}\label{ele}
    L(t)=\prod_i\lambda_i(t)\,.
\end{equation}
By convention we consider $\{\lambda_i(t)\}$ to be ordered at any time $t$ with $\lambda_0(t)$ the eigenvalue with the smallest magnitude. Similarly when resolving by momentum $k$, $\lambda_{0,k}(t)$ is the lowest magnitude eigenvalue with momentum $k$. For the return rate derivative one finds
\begin{equation}
\dot{l}(t)=-\frac{1}{N}\tr\dot{\bm M}(t){\bm M}^{-1}(t)\,.
\end{equation}
We will also use $\lambda_0(t)$ or $\lambda_{0,k}(t)$ as a proxy for the Fisher zeroes by studying when they become sufficiently close to zero in a finite size system. The beginning and ends of regions of time for which eigenvalues are zero give at leats some of the critical times. We note that a divergence in the density of eigenvalues inside a critical region would also in principle give a critical time, but we can not track these.

The bulk contribution can be found from the reciprocal space form, where larger system sizes are possible. For open systems we focus on a cylindrical geometry with on\textcolor{red}{e} periodic and one open direction, where we have access to a parallel wavevector to resolve the Loschmidt eigenvalues. We will further consider both armchair and zig-zag edges for the open boundary case and compare them.

Together the Chern number and bulk boundary correspondence~\cite{Teo2010} predicts the number of protected bands crossing the gap. However it does not dictate how they cross, which can influence whether they describe Majorana fermions and have Majorana zero modes at zero energy~\cite{Sedlmayr2017,Glodzik2023}. For the cylindrical geometry only topological bands which cross at time reversal invariant momenta have Majorana zero modes at zero energy, these can contribute an extra term to the boundary return rate~\cite{Maslowski2026a} and the number of these present can depend on the orientation of the edges~\cite{Sedlmayr2017,Glodzik2023}. However the Majorana zero modes are rather artifacts of the special cylindrical geometry where there are two distinct edges, and would not exist for a physical two dimensional system with a single edge. Due to this and to the difficulty of extracting the boundary return rate from a finite size scaling analysis for the model we study here, we will not analyse these contributions in this article.

\section{Model}\label{sec:mods}

\begin{figure}[t]
\includegraphics[height=0.85\columnwidth]{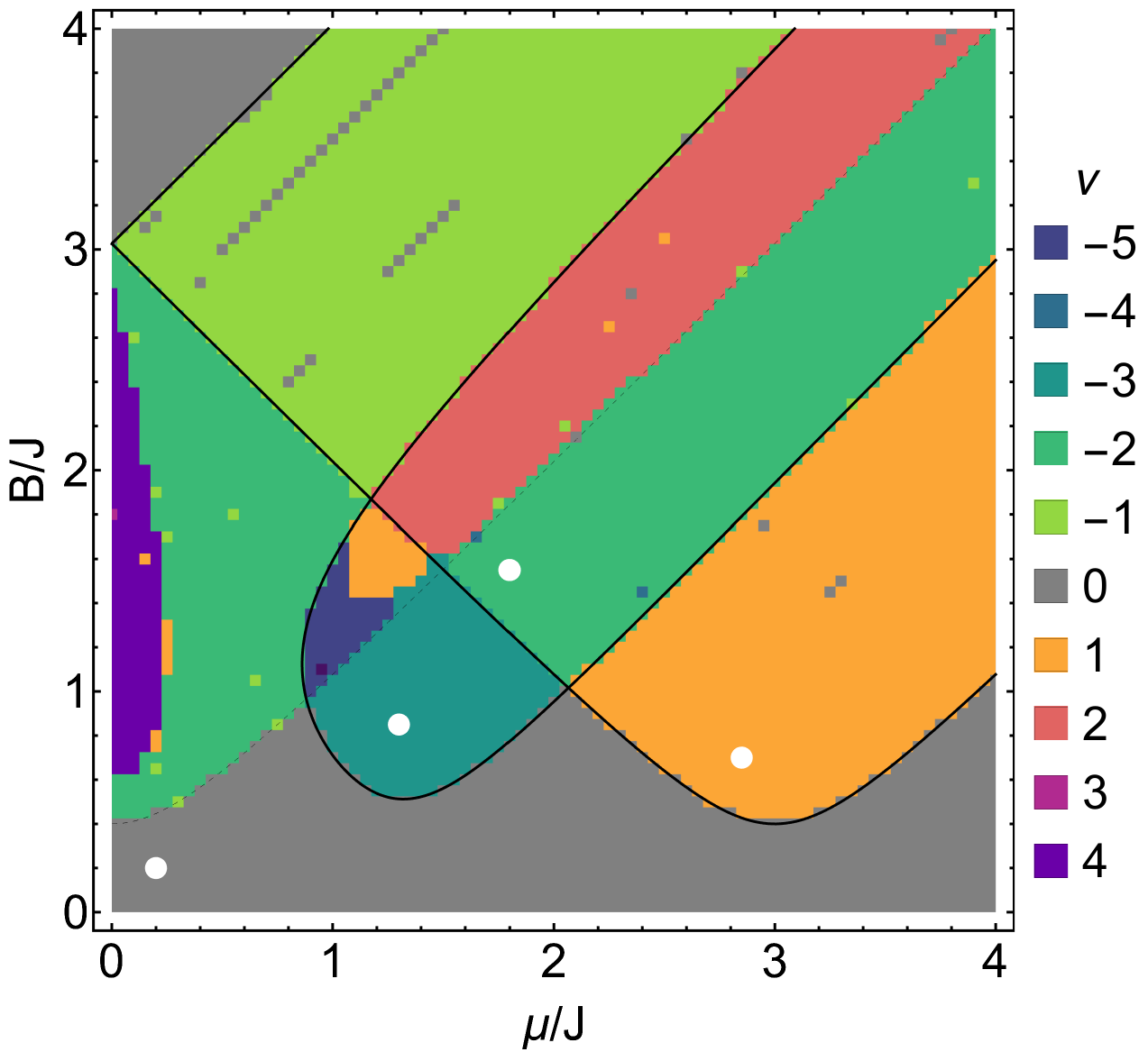}
\caption{The topological phase diagram for the topological superconductor~\cite{Sedlmayr2017}, Eq.~\eqref{graphene}, with $\alpha=0.4J$ and $\Delta=0.4J$. White circles are the points in the phase diagram between which we quench, see table \ref{tab:quenches}.} 
\label{fig:phase_hex}
\end{figure}

The model we use as our test bed is a two dimensional topological superconductor defined on a hexagonal lattice~\cite{Sedlmayr2017}. Hexagonal lattice models with proximity induced superconductivity and spin orbit coupling have been studied extensively, and prove to have rich topological phase diagrams~\cite{Black-Schaffer2014,Dutreix2014,Sedlmayr2015,Wang2016,Dutreix2017,Sedlmayr2017,Kaladzhyan2017,Pangburn2023a,Crepieux2023,Pangburn2023}, making it a good candidate for investigating a wide range of quenches between different topological phases. This particular model has a complex phase diagram with a wider range of available Chern numbers $\nu\in\{-5,-4,\ldots,5\}$, see Fig.~\ref{fig:phase_hex}. Additionally the hexagonal lattice means that there are different types of edges along which we can probe the boundary contributions following the quenches.

The model includes Rashba spin-orbit coupling, $\alpha$, a proximity induced $s$-wave superconductivity, $\Delta$, and a magnetic Zeeman potential, $B$. We define $J$ as the strength of the nearest-neighbour hopping and $\mu$ as the chemical potential. The Hamiltonian is~\cite{Sedlmayr2017}:
\begin{align}\label{graphene}
\hat{H}=&-\sum_{<i,j>}\Psi^\dagger_{i}\left[J{\bm\tau}^z+i\alpha{\bm\tau}^z(\vec{\delta}_{ij}\times{\vec {\bm\sigma}})_z\right]\Psi_{j}\nonumber\\
&+\sum_{j}\Psi^\dagger_{j}\left(B{\bm\sigma}^z-\mu{\bm\tau}^z+\Delta{\bm\tau}^x\right)\Psi_{j}\,.
\end{align}
where $\Psi^\dagger_{j}=\{c^\dagger_{j\uparrow},c^\dagger_{j\downarrow},c_{j\downarrow},-c_{j\uparrow}\}$ with $c_{ j}^{(\dagger)}$ annihilating (creating) a particle of spin $\sigma$ at a site labelled by $j$, and $\vec\delta_{ij}$ is the nearest neighbour vector between sites $i$ and $j$. Here ${\bm\tau}^j$ and ${\bm\sigma}^j$ are Pauli matrices for the particle-hole and spin subspaces respectively. In the standard way everywhere a tensor product is implied between the ${\bm\tau}^j$ and ${\bm\sigma}^j$ matrices, and identity matrices are not explicitly written. 

After a Fourier transform the Hamiltonian can be written as $H=\sum_{\vec k}\Psi^\dagger_{\vec k}\mathcal{H}(\vec k)\Psi_{\vec k}$ with 
\begin{equation}
\Psi_{\vec{k}}=(\hat{a}_{{\vec{k}},\uparrow},\hat{b}_{{\vec{k}},\uparrow},\hat{a}_{{\vec{k}},\downarrow},\hat{b}_{{\vec{k}},\downarrow},\hat{a}^{\dag}_{{\vec{k}},\downarrow},\hat{b}^{\dag}_{{\vec{k}},\downarrow},-\hat{a}^{\dag}_{{\vec{k}},\uparrow},-\hat{b}^{\dag}_{{\vec{k}},\uparrow})^T\,.
\end{equation}
Here $\hat{a}^{(\dagger)}_{{\vec{k}},\sigma}$ and $\hat{b}^{(\dagger)}_{{\vec{k}},\sigma}$ refer to the sublattices of the hexagonal lattice in which $\mu$, $B$ and $\Delta$ are all diagonal. Momentum, or equivalently  wavevector as we set the lattice spacing between nearest neighbours as $a=1$ and also use $\hbar=1$, is $\vec{k}=(k_x,k_y)$. $k_x$ is a zig-zag direction and $k_y$ armchair. The Hamiltonian density is, with the convention $\tau\otimes\sigma\otimes\textrm{sublattice}$,
\begin{equation}\label{kham}
    \mathcal{H}(\vec k)=
    \begin{pmatrix}
        {\bm f}_{\vec k}  & 0\\
        0  & -{\bm f}^\dagger_{-\vec k} 
    \end{pmatrix}-\mu{\bm\tau}^z+B{\bm\sigma}^z+\Delta{\bm\tau}^x\,,
\end{equation}
where
\begin{align}
    {\bm f}(\vec k)=&
    -\sum_{\{\vec\delta\}}
    e^{i \vec k\cdot\vec\delta}
    \begin{pmatrix}
        0 & J &  0 & -\alpha\vec{\delta}\cdot(1,i)\\
        0& 0  &  0 & 0\\
        0 &  \alpha \vec{\delta}\cdot(1,-i) & 0  & J\\
        0 & 0 & 0 & 0 
    \end{pmatrix}
    \nonumber\\&
    -\sum_{\{\vec\delta\}}
    e^{-i \vec k\cdot\vec\delta}
    \begin{pmatrix}
        0 & 0 & 0 & 0 \\
        J & 0 & \alpha\vec{\delta}\cdot(1,i) & 0\\
        0 & 0 & 0  & 0 \\
        \alpha\vec{\delta}\cdot(-1,i) & 0 & J & 0 
    \end{pmatrix}\,.
\end{align}
The set of nearest neighbour vectors between the sublattices is
\begin{equation}\label{deltavector}
\{\vec \delta\}=\left\{\left(\frac{\sqrt{3}}{2},-\frac{1}{2}\right),\left(-\frac{\sqrt{3}}{2},-\frac{1}{2}\right),\left(0,1\right)\right\}
\end{equation}
for the hexagonal lattice.

This model possesses only a particle-hole symmetry $\mathcal{C}$:
\begin{equation}
	\mathcal{C}\hc_{{\vec{k}}}\mathcal{C}=-\hc^*_{-{\vec{k}}}.
\end{equation}
where $\mathcal{C}^2=1$, placing it in the D class of the topological periodic table with $\mathbb{Z}$ invariants in two dimensions~\cite{Ryu2010,Chiu2016}. The appropriate invariant is the Chern number, or equivalently the Thouless-Kohmoto-Nightingale-den Nijs (TKNN) invariant~\cite{Thouless1982}. The Chern number can be conveniently written as ~\cite{Girvin2019}
\begin{equation}
\nu=\int \frac{d^2k}{2\pi} \sum_ni\langle\partial_{\vec{k}}u_{n\vec{k}}|\times|\partial_{\vec{k}}u_{nk}\rangle
\end{equation}
for a Bloch band $u_{n\vec{k}}$, which is equivalent to
\begin{equation}\label{cherneq}
\nu=\int \frac{d^2k}{2\pi} \sum_{n \neq\ell}\mathcal{H}^{x}_{n \ell}\mathcal{H}^{y}_{\ell n}\frac{\textrm{sgn}\,[\epsilon_n]\Theta[-\epsilon_n\epsilon_\ell]}{(\epsilon_n-\epsilon_\ell)^2}\,.
\end{equation}
Here $S$ is the rotation to the energy eigenbasis of the Hamiltonian, \emph{i.e.}~$S^{-1}\hc_{{\vec{k}}}S$ is diagonal with eigenvalues $\epsilon_n$ and we defined $\hc^{x,y}\equiv S^{-1}\partial_{k_{x,y}}\hc_{{\vec{k}}}S$. Using Eq.~\eqref{cherneq} the phase diagram can be straightforwardly calculated numerically, see Fig.~\ref{fig:phase_hex}, on which we also mark the quenches we consider. The bandstructures of these particular points in the phase diagram can be seen in Figs \ref{fig:bs_hex}.

As there are too many possible quenches to show every example we focus here on some exemplary cases for which good data could be found. An extensive overview and more data can be found at~\onlinecite{Maslowski2026}. We first set $\alpha=0.4J$ and $\Delta=0.4J$ to get a reasonable gap size and interesting phase diagram. We then found the points in this particular phase diagram with the largest gaps for each phase. Next, we considered the DQPTs for quenches between all these points, selecting some for presentation here, see table \ref{tab:quenches} for parameters used in the various quenches in this article.

\begin{table}
    \begin{center}
    \begin{tabular}{|c|c|}
        \hline \multicolumn{2}{|c|}{HTS Model} \\
        \hline Parameters & Chern No.\\
        $(\mu/J,B/J)$ & $\nu$  \\\hline
        $(0.2,0.2)$ & 0   \\\hline
        $(2.85,0.7)$ & 1 \\\hline
        $(1.8, 1.55)$ & -2 \\\hline
        $(1.3,0.85)$ & -3  \\\hline
    \end{tabular}
    \caption{The points in the phase diagrams used for the quenches, see Fig.~\ref{fig:phase_hex}. In all cases $\alpha=0.4J$ and $\Delta=0.4J$.}
    \label{tab:quenches}
    \end{center}
\end{table}

\section{Results for The Dynamical Bulk Boundary Correspondence}\label{sec:hexdbb}

We now turn to results for the DBBC. Examples of the DQPTs can be found in appendix \ref{app:bulkdqpts} and Fig.~\ref{fig:rrh1}, where discontinuities in the return rate derivative can be clearly seen at critical times. As we are not able to directly calculate the critical times they are inferred from the behaviour of the lowest Loschmidt eigenvalue $\lambda_0(t)$ calculated for a lattice system of size $N=200^2$ with periodic boundary conditions. When $\lambda_0(t)$ is zero the system is in a critical region, the onset and end of which are the critical times. We note that further critical times can appear inside the critical region~\cite{Maslowski2024b}.

\begin{figure}
\includegraphics[height=0.45\columnwidth]{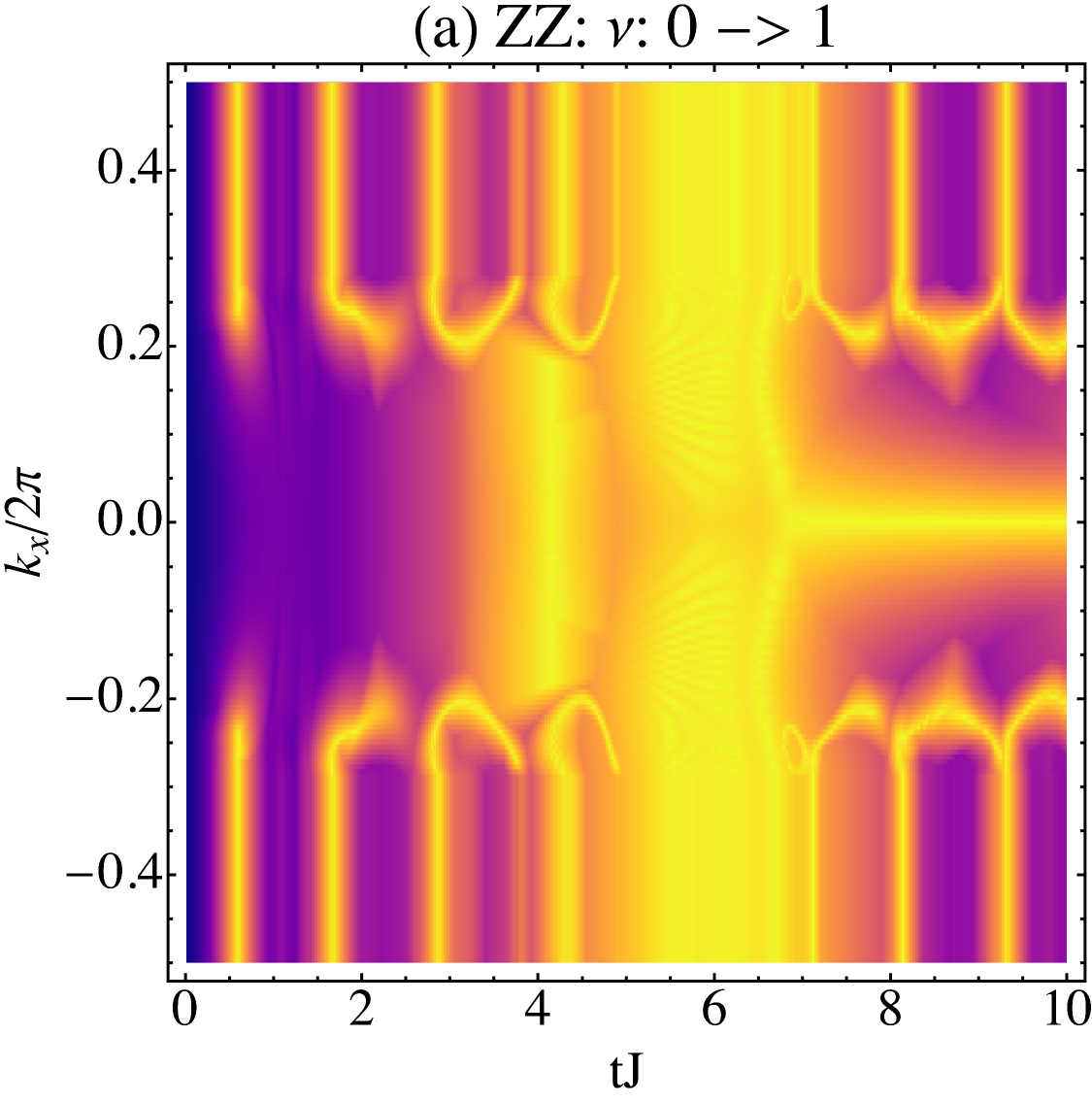}
\includegraphics[height=0.45\columnwidth]{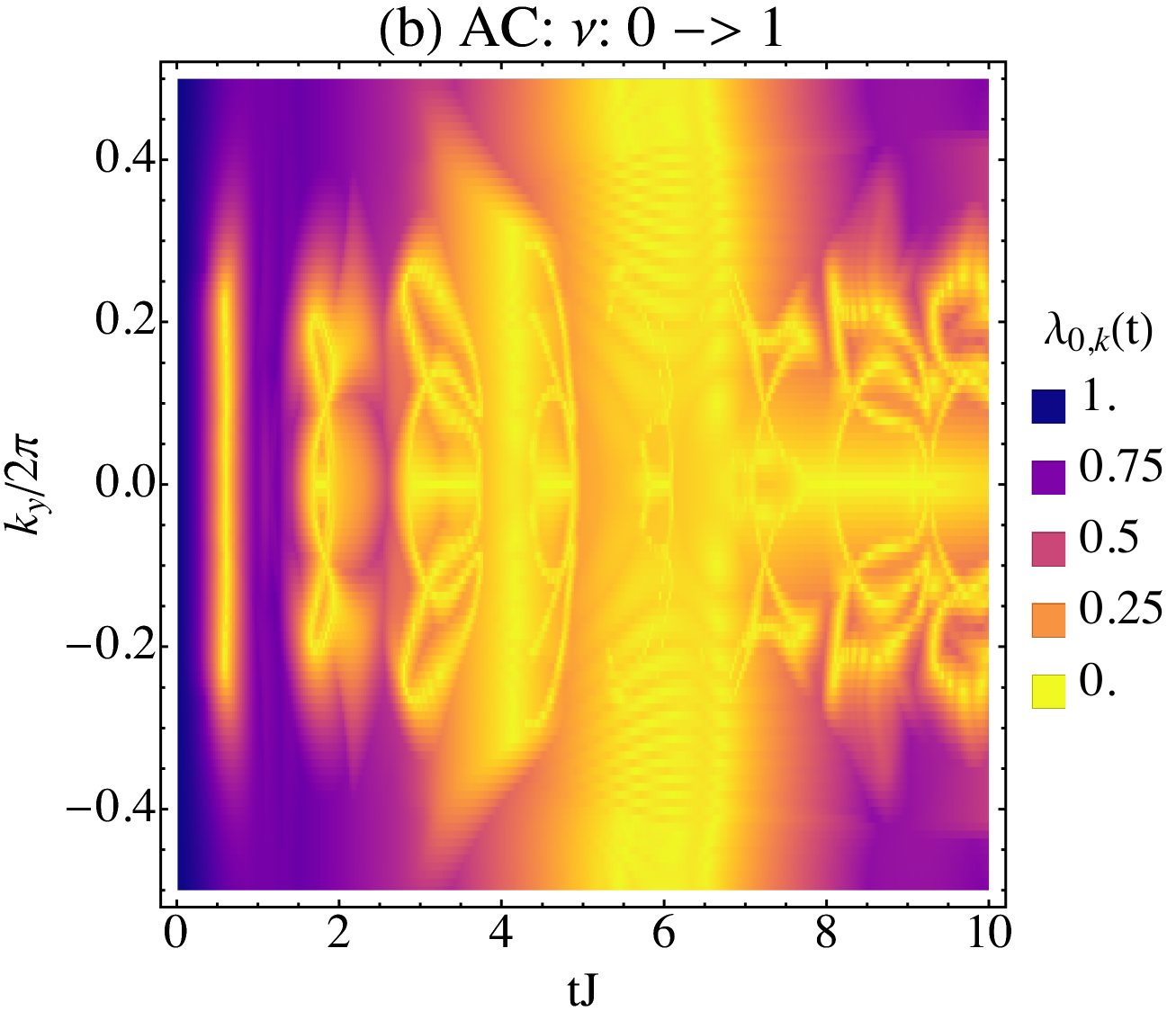}\\
\includegraphics[width=0.8\columnwidth]{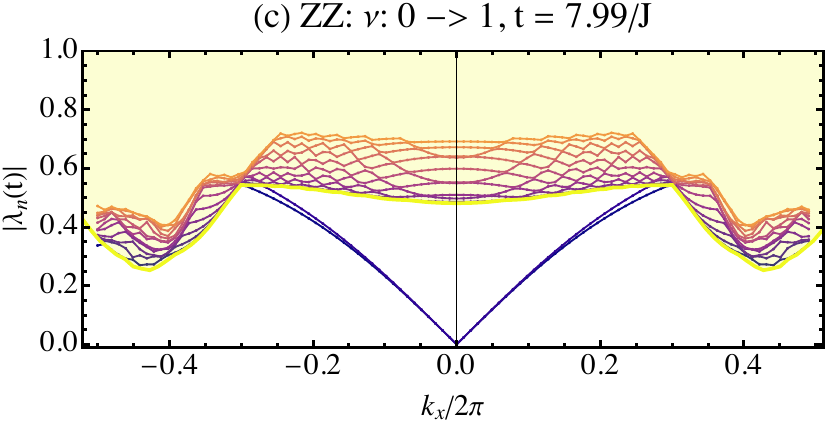}
\caption{In panels (a,b) the lowest Loschmidt eigenvalue $\lambda_{0,k}(t)$ as a function of momentum and time is plotted. This is calculated for a system of size $102\times100$ unit cells in the open and periodic directions respectively for both zig-zag (ZZ, panel a) and armchair (AC, panel b) edges.  Panel (c) shows a cut at a particular time for the zig-zag edges with the lowest 12 eigenvalues $\lambda_{n,k}(t)$ and the lowest bulk eigenvalue calculated for periodic boundary conditions ($N=200^2$) giving the limit of the shaded region. In (a,b) the bulk critical regions appear as vertically orientated curved bright lines, and the lowest eigenvalue of the band of eigenvalues crossing the gap appear as horizontally orientated bright lines.  See table \ref{tab:quenches} for more details. In panel (c) the in-gap modes are clearly visible, which correspond to the bright horizontal line in (a) and is also visible for the armchair case (b). See the main text for further discussion.} 
\label{fig:evrdehex}
\end{figure}

We generally focus on cases where the critical regions of zeroes are short. This has several advantages: consecutive critical regions do not overlap over the time periods considered, and in-gap modes give signatures in $\lambda_{0,k}(t)$ easily distinguished from the bulk behaviour. Fig.~\ref{fig:evrdehex} shows the simplest example where we expect in-gap modes due to the DBBC, a quench $\nu:0\to1$. For both zig-zag and armchair edges in-gap modes can be seen to form, however they only occur if the gap closes at the correct momentum and so not all critical times are associated with the formation of in-gap modes. This is not however the same for the different types of boundary. One can see that after projection onto the parallel momentum the armchair example has many more gap-closings at $k_y=0$ which lead to boundary contributions which have no counterpart for the zig-zag edged system. IT is unclear if the in-gap mods are associated with a form of weak topology (one which depends on the direction of the edges) or if they are of topological origin at all. We note that the opposite quench shows no in-gap modes in the spectrum of $\lambda_n(t)$ forming at all, as would be expected from the DBBC.

\begin{figure}
\includegraphics[height=0.45\columnwidth]{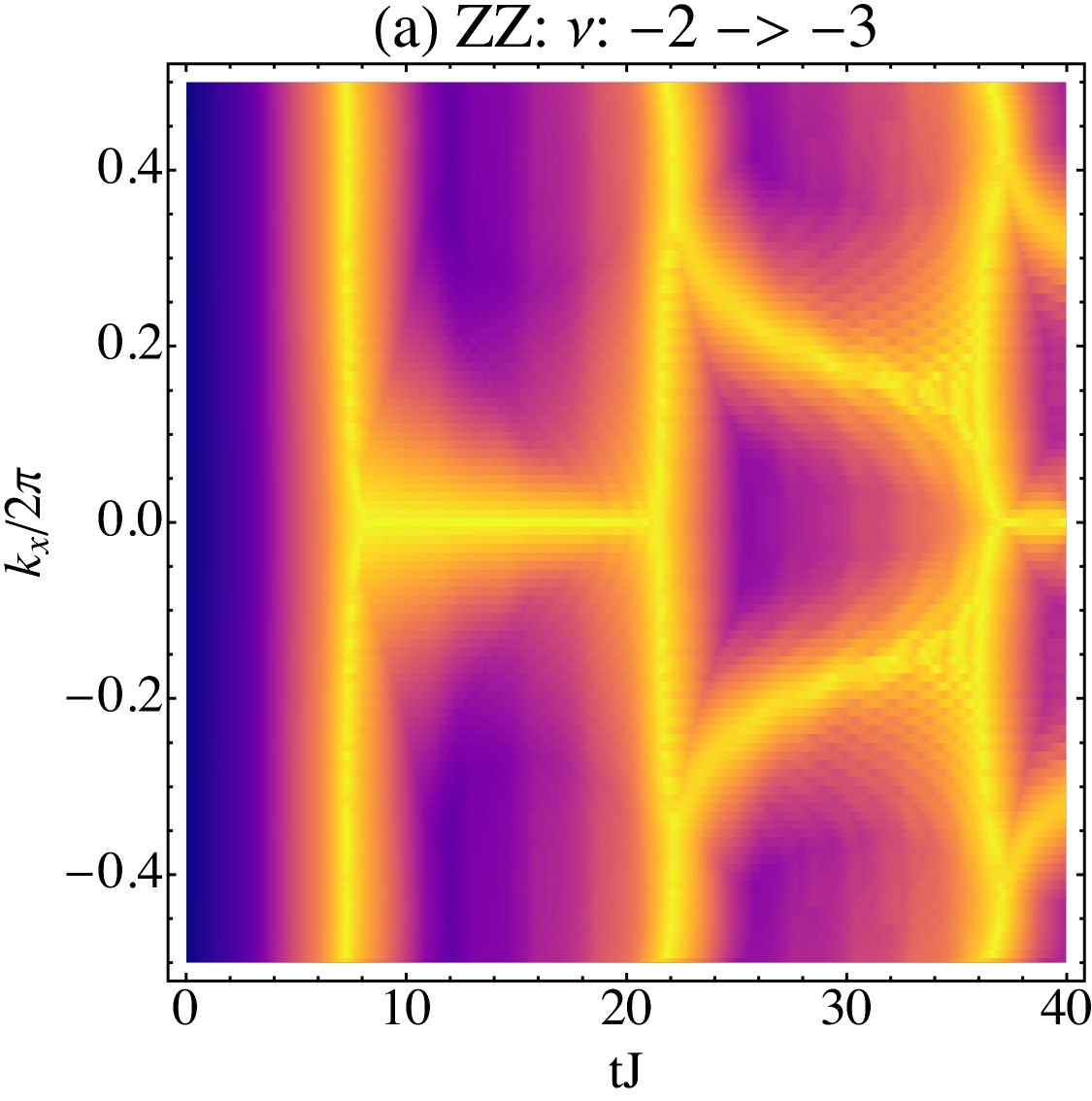}
\includegraphics[height=0.45\columnwidth]{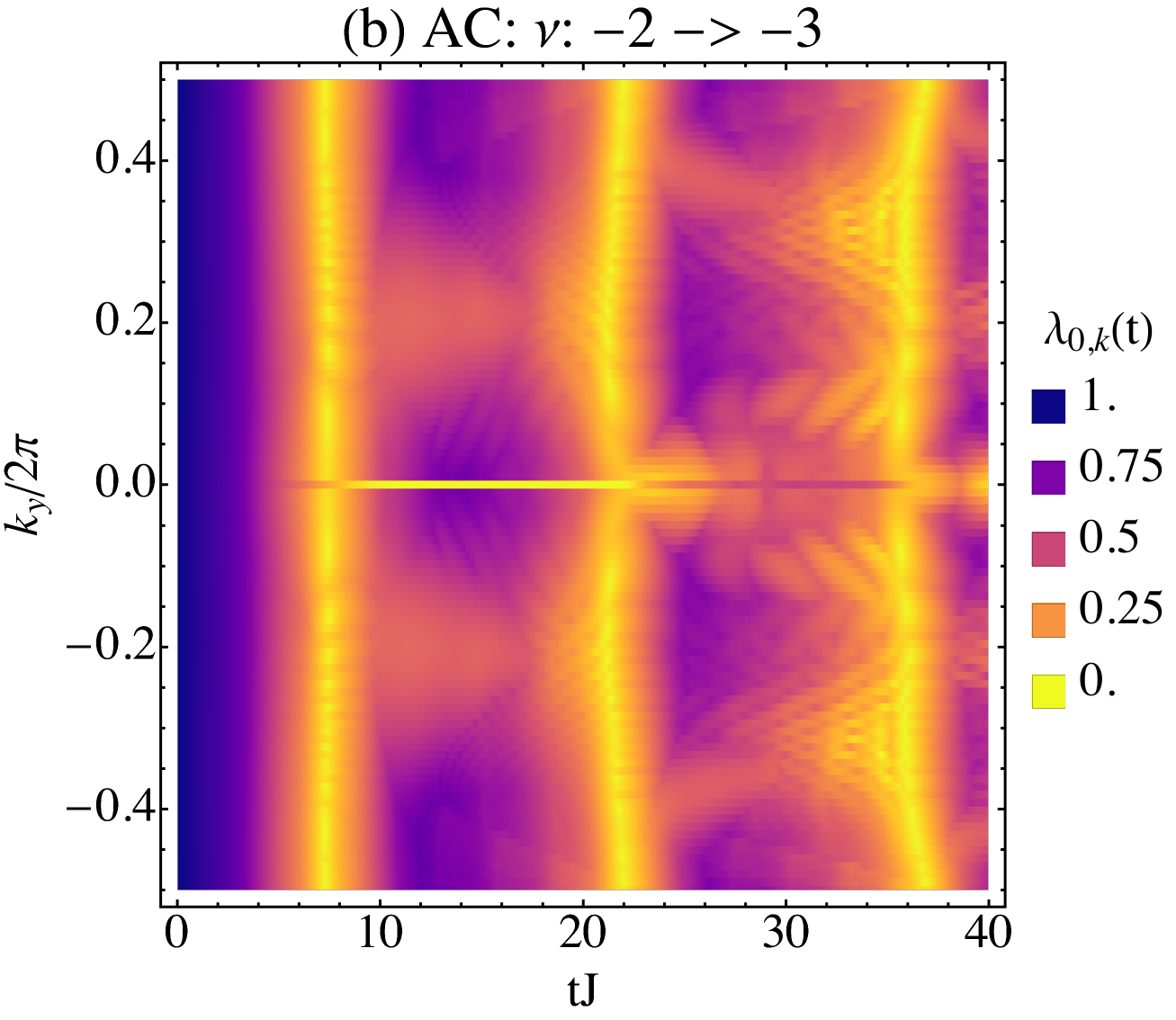}\\
\includegraphics[width=0.8\columnwidth]{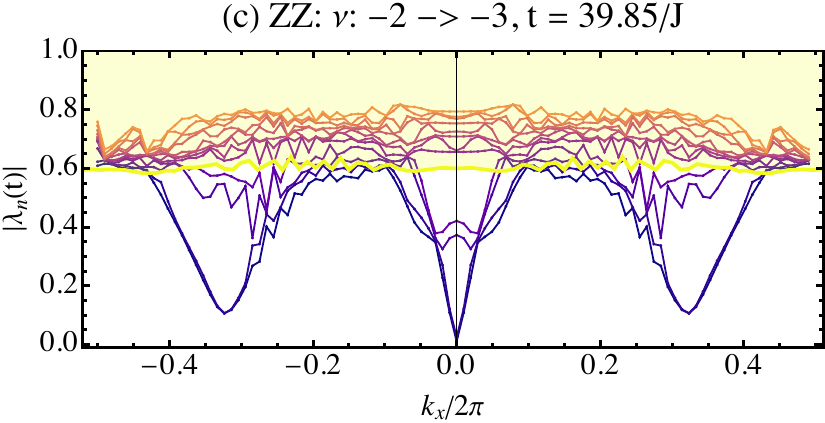}
\caption{In panels (a,b) the lowest Loschmidt eigenvalue $\lambda_{0,k}(t)$ as a function of momentum and time is plotted for both zig-zag (a) and armchair (b) edges.  Panel (c) shows a cut at a particular time for the zig-zag edges with the lowest 12 eigenvalues $\lambda_{n,k}(t)$ and the lowest bulk eigenvalue. See Fig.~\ref{fig:evrdehex} and table \ref{tab:quenches} for more details. See the main text for a discussion of the results.} 
\label{fig:evrdehex2}
\end{figure}

In Fig.~\ref{fig:evrdehex2} examples are shown for $\nu:-2\to-3$. In such a case it is unclear what to expect from a naive version of the DBBC~\cite{Sedlmayr2018,Maslowski2026a}, although if the topological phase of the time evolving Hamiltonian is a crucial ingredient we may expect in-gap modes to occur for this case. This is clearly visible in Fig.~\ref{fig:evrdehex2} for both zig-zag and armchair edges, with no essential difference in the appearance of the in-gap modes visible for the two cases. As can be seen most clearly in panels (a,c) for the zig-zag edges additional in-gap modes can also appear away from $k=0$ though they do not touch $|\lambda_{0,k}(t)|\to0$, though it is feasible this may be a finite size effect. From Fig.~\ref{fig:bs_hex}(c,d) we note that in the equilibrium bandstructures for the $\nu=-3$ phase both zig-zag and armchair edges have crossings away from $k=0$, however no quantitative relation to the Loschmidt in-gap modes is evident.

\begin{figure}
\includegraphics[height=0.45\columnwidth]{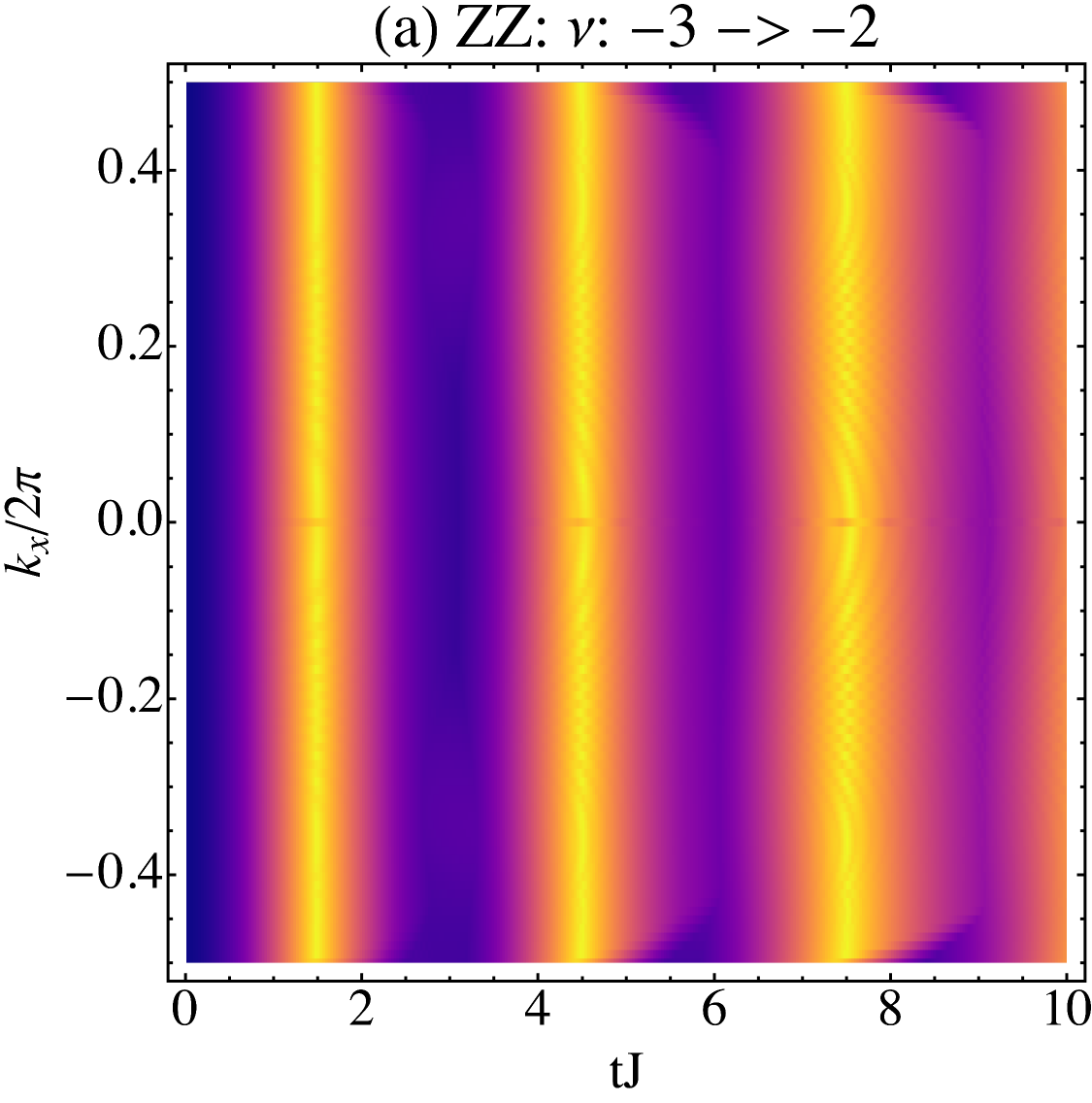}
\includegraphics[height=0.45\columnwidth]{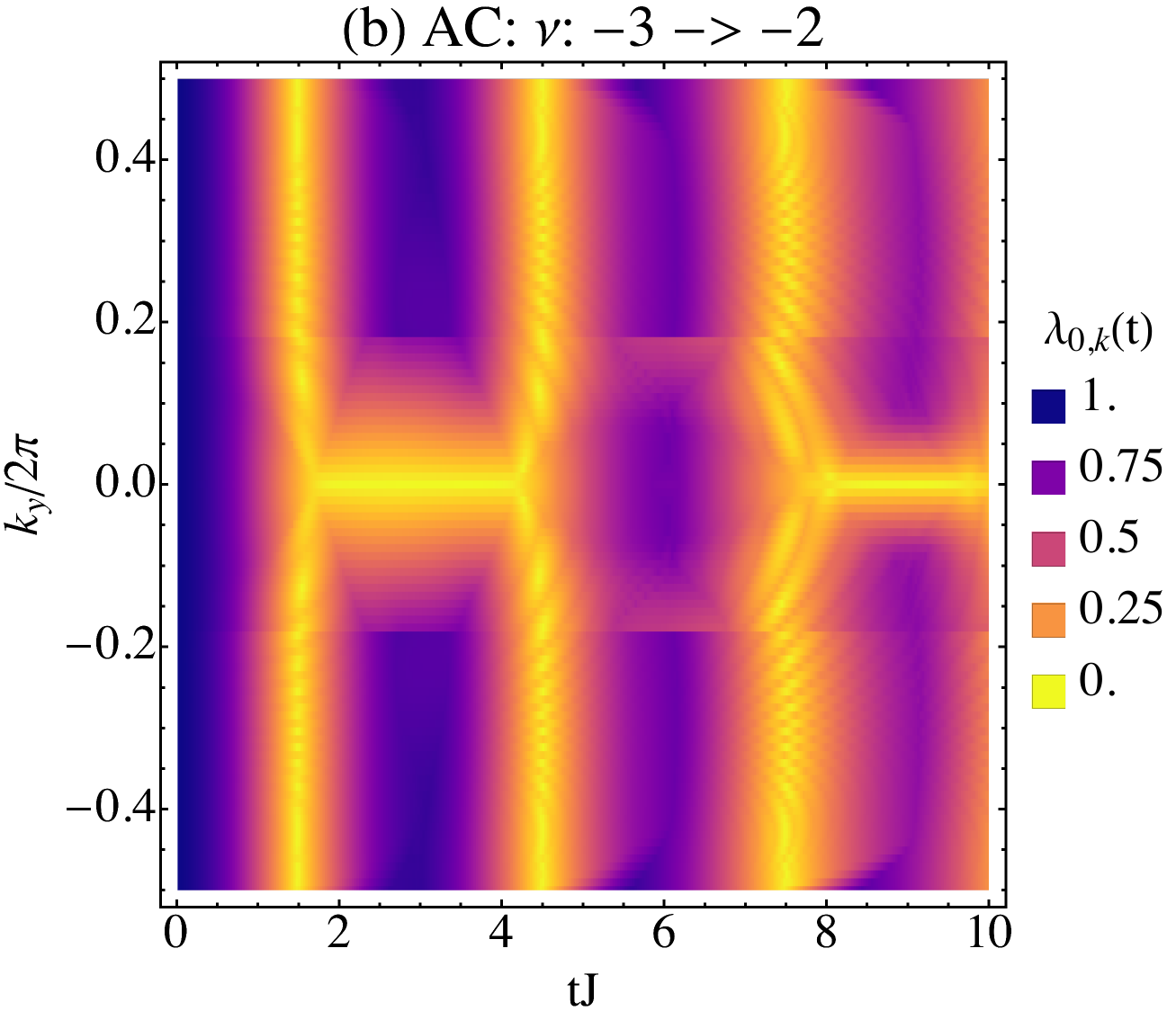}\\
\includegraphics[width=0.8\columnwidth]{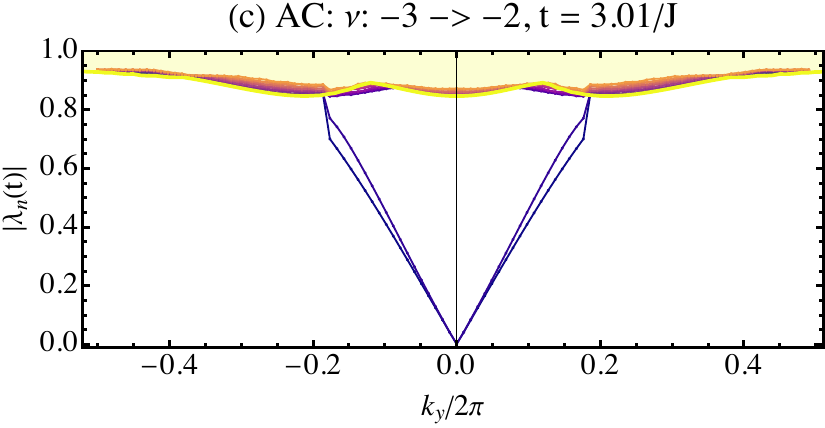}
\caption{In panels (a,b) the lowest Loschmidt eigenvalue $\lambda_{0,k}(t)$ as a function of momentum and time is plotted for both zig-zag (a) and armchair (b) edges.  Panel (c) shows a cut at a particular time for the armchair edges with the lowest 12 eigenvalues $\lambda_{n,k}(t)$ and the lowest bulk eigenvalue. See Fig.~\ref{fig:evrdehex} and table \ref{tab:quenches} for more details. See the main text for a discussion of the results.} 
\label{fig:evrdehex3}
\end{figure}

Let us now turn to the opposite quench, $\nu:-3\to-2$ in Fig.~\ref{fig:evrdehex3}. As is clear from a comparison of panels (a) and (b), only the armchair edge shows any in-gap modes forming with the DBBC clearly demonstrated in that case. It is interesting to compare this dynamical behaviour to the equilibrium bandstructures in Fig.~\ref{fig:bs_hex}. For the zig-zag case $\nu=-3$ has a single crossing at $k=0$ and the $\nu=-2$ phase has no crossing of boundary modes at $k=0$, they occur elsewhere in the Brillouin zone. By contrast for the armchair case $\nu=-3$ still has a single crossing at $k=0$ but the $\nu=-2$ phase has two crossings of boundary modes at $k=0$. Thus if we take only these crossings into account the zigzag case would be effectively $1\to0$ and the armchair case $1\to2$. This is purely observational, and it is not clear if it is followed in all cases\textcolor{red}{\cancel{,}.} Indeed we see that $\nu:1\to-2$ has the same behaviour of equilibrium band crossings for the armchair edges, but no in-gap Loschmidt modes form during time evolution~\cite{Maslowski2026}. Although the nature of the equilibrium phases must be important for the DQPTS and DBBC, no simple relation between equilibrium and non-equilibrium behaviour is found.

Typically we can observe that if $H_1$ belongs to an odd Chern number phase then the DBBC is most likely present, but if the Chern number is even no clear statement can be made. Furthermore the effect may depend on orientation of the edges. We see clear in-gap modes which cross at $k=0$ ($k=\pi$ would presumably also be possible, but we do not look at such equilibrium phases here), and other modes inside the gap which do not touch zero. To what extent each of these are topological, and to what extent weak topology may play a role remains unclear. A reasonable supposition is that the non-Hermitian Loschmidt matrix is responsible for the behaviour, but we have no current clear theory, and two dimensional non-Hermitian topology remains under-explored~\cite{Sirker2026}.

\section{Conclusions}\label{sec:con}

In conclusion, although a natural two dimensional analogue of the dynamical bulk-boundary correspondence does continue to hold in some form for the model considered here, we see that its appearance is not exactly in correspondence with a naive counting of the Chern number for the initial and time-evolving Hamiltonians. At first sight it depends only on the parity of the Chern number, but this does not hold perfectly. This suggests that this phenomenon, like DQPTs themselves, can not be completely determined by a mapping to the equilibrium phase diagrams, but contains non-trivial physics of the dynamics. To investigate further would require investigations on a wider variety of models.

Some caution is required however due to finite size effects potentially obscuring effects. Our results also suggests caution is required in relying too much on the most simple models when investigating dynamical phenomena. The origin of this dynamical boundary effect and its exact relation to the bulk properties remains unexplained, and is an important question for future work. This article demonstrates the need for a more fundamental understanding of the DBBC, and the dynamical topology involved, but further investigations, including the effects disorder and possible topological invariants, are needed.

With regards to possible measurements of the effects we note that although the Loschmidt echo can be measured in several set ups, as discussed in our introduction, this particular realisation would be very complicated to achieve. Not only is the model two dimensional with a relatively large subspace, to see boundary effects would require finite size scaling of the experiment. A more promising direction is to find correlates and proxies of both the DQPTs and DBBC in more easily accessible physical observables, this is an ongoing area of research.

\acknowledgments
N.S.~acknowledges support from the National Science Centre (NCN, Poland) under the grant 2024/53/B/ST3/02600. All data used in this article, as well as additional data sets and figures can be found on Zenodo~\cite{Maslowski2026}.

\appendix

\section{Examples of the Bandstructures}\label{app:bands}

\begin{figure}
\includegraphics[width=0.45\columnwidth]{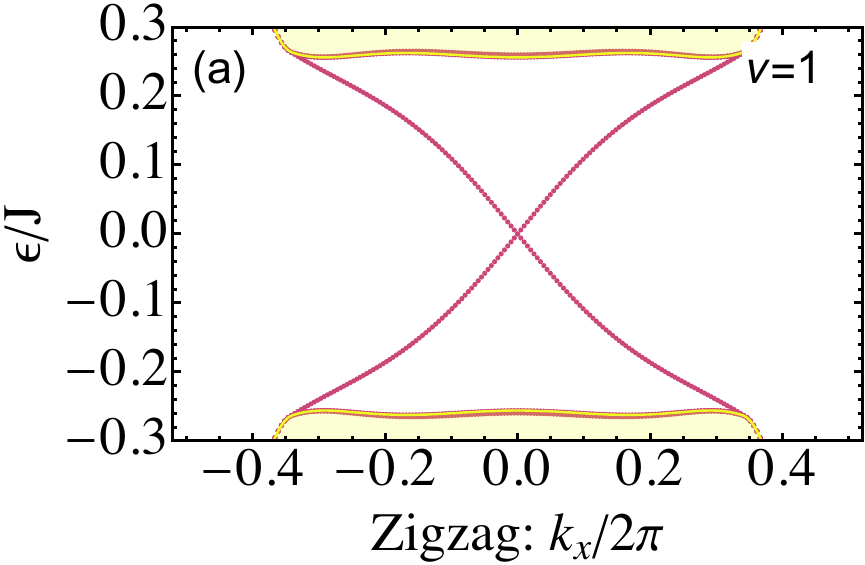}
\includegraphics[width=0.45\columnwidth]{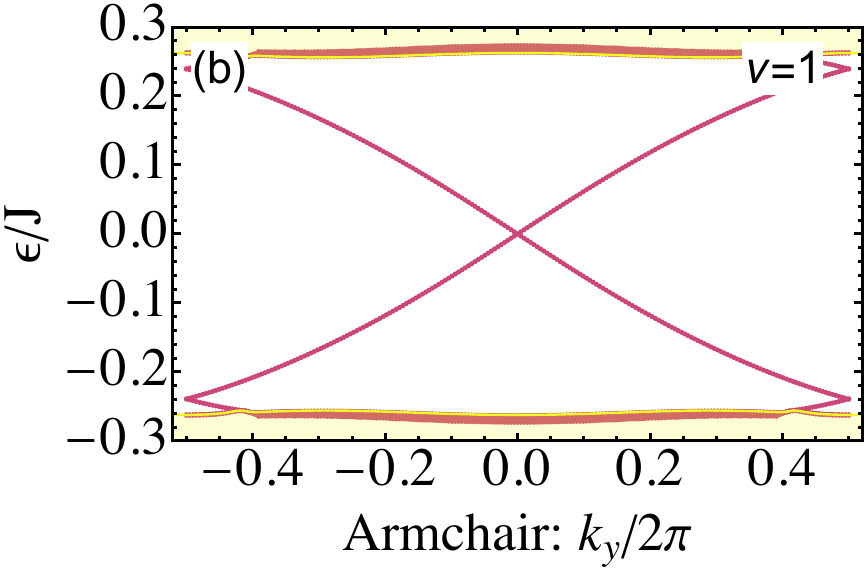}\\
\includegraphics[width=0.45\columnwidth]{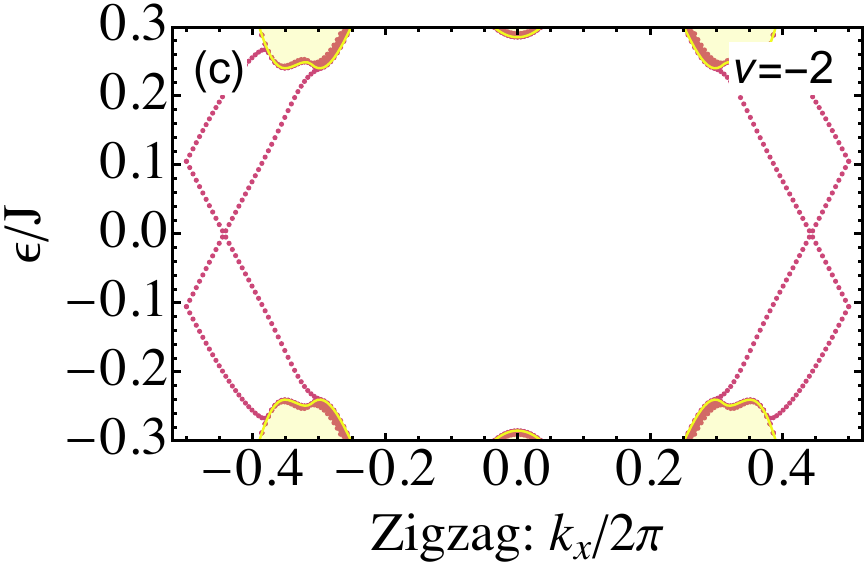}
\includegraphics[width=0.45\columnwidth]{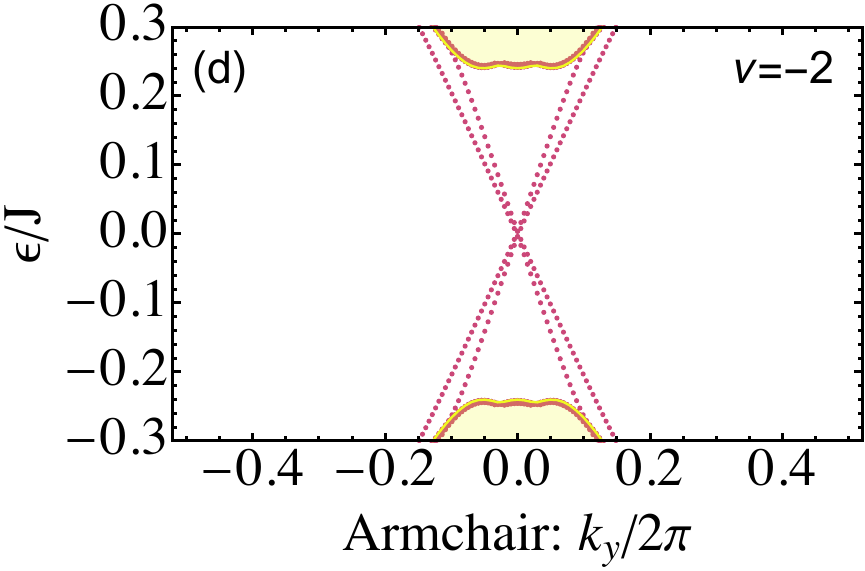}\\
\includegraphics[width=0.45\columnwidth]{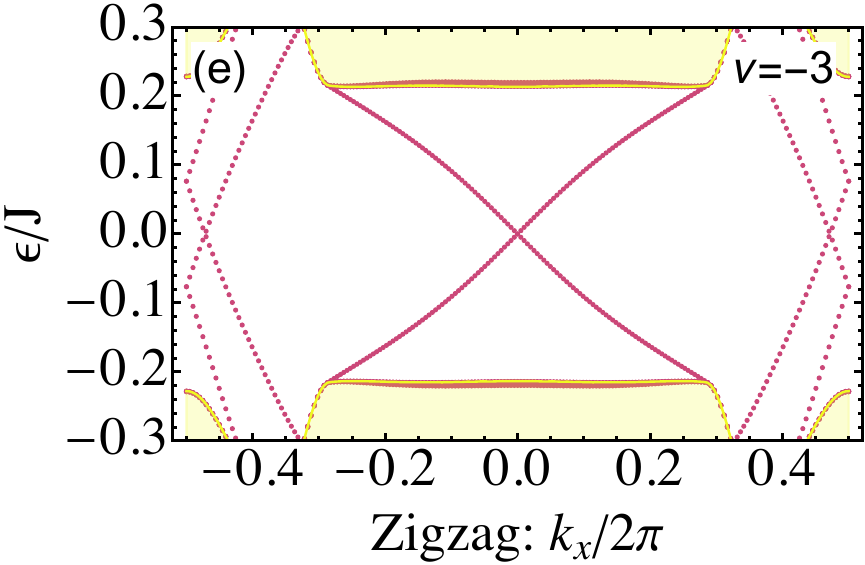}
\includegraphics[width=0.45\columnwidth]{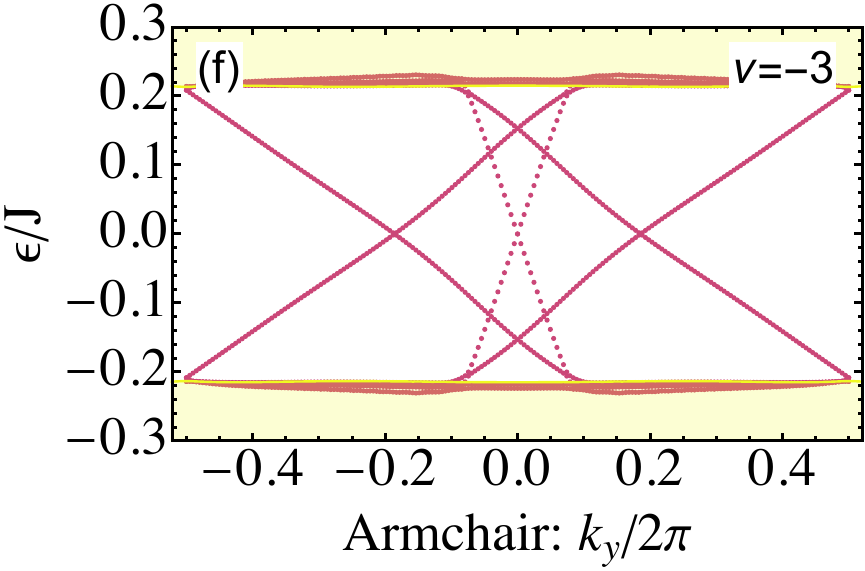}
\caption{Bandstructure examples for the model, \eqref{graphene}, where $\alpha=0.4J$ and $\Delta=0.4J$ and further parameters are as in table \ref{tab:quenches}. The topologically protected bands can be seen crossing the gap, Chern numbers $\nu$ are as marked on the figures. Note as this is a chiral system left moving and right moving bands are on opposite edges of the system.} 
\label{fig:bs_hex}
\end{figure}

In Fig.~\ref{fig:bs_hex} we show the bandstructures projected along both the zig-zag and armchair directions for the exemplary points in the phase diagrams with $\nu\in\{1,-2,-3\}$ between which we quench. See the Hamiltonian \eqref{graphene}, Fig.~\ref{fig:phase_hex}, and table \ref{tab:quenches}. As can be clearly seen the crossings can occur in different ways for the two different types of edges, which has an effect on the behaviour of $\lambda_{0,k}(t)$. The protected number of band crossings is of course given by the Chern number $\nu$.

\section{Examples of the Dynamical Quantum Phase Transitions}\label{app:bulkdqpts}

\begin{figure*}
\includegraphics[width=0.95\columnwidth]{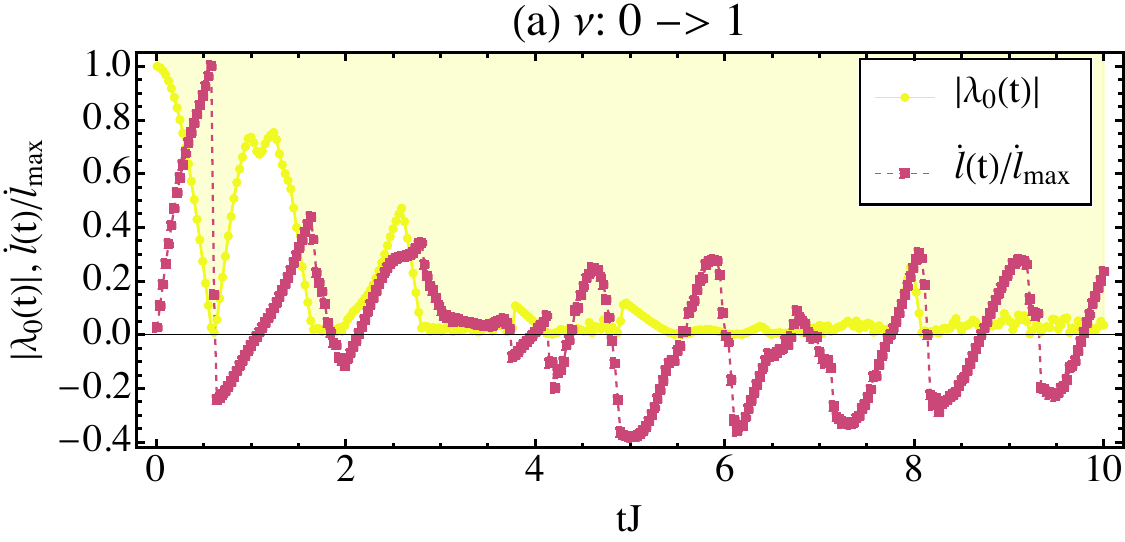}
\includegraphics[width=0.95\columnwidth]{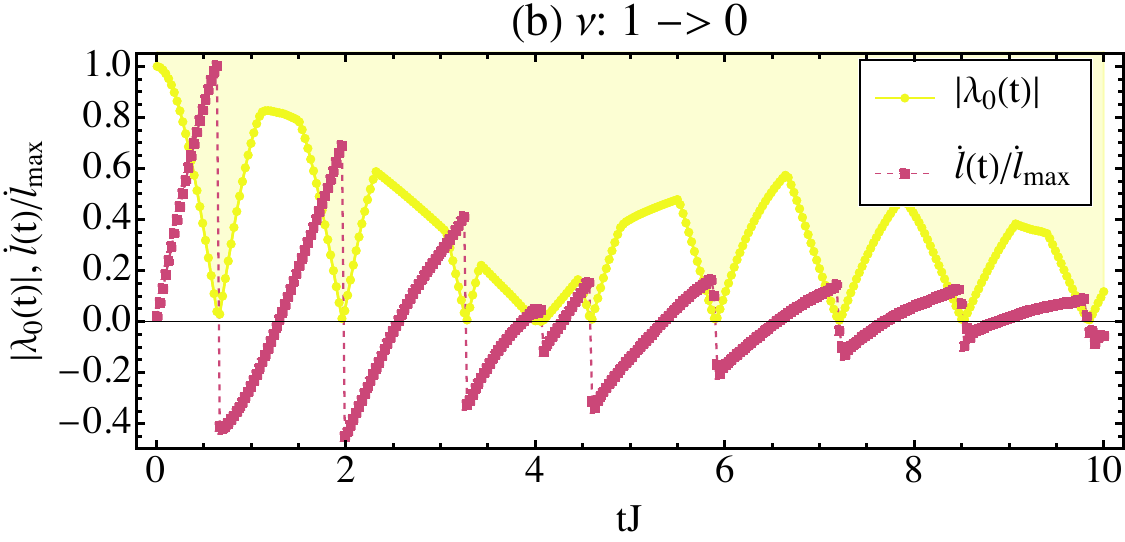}\\
\includegraphics[width=0.95\columnwidth]{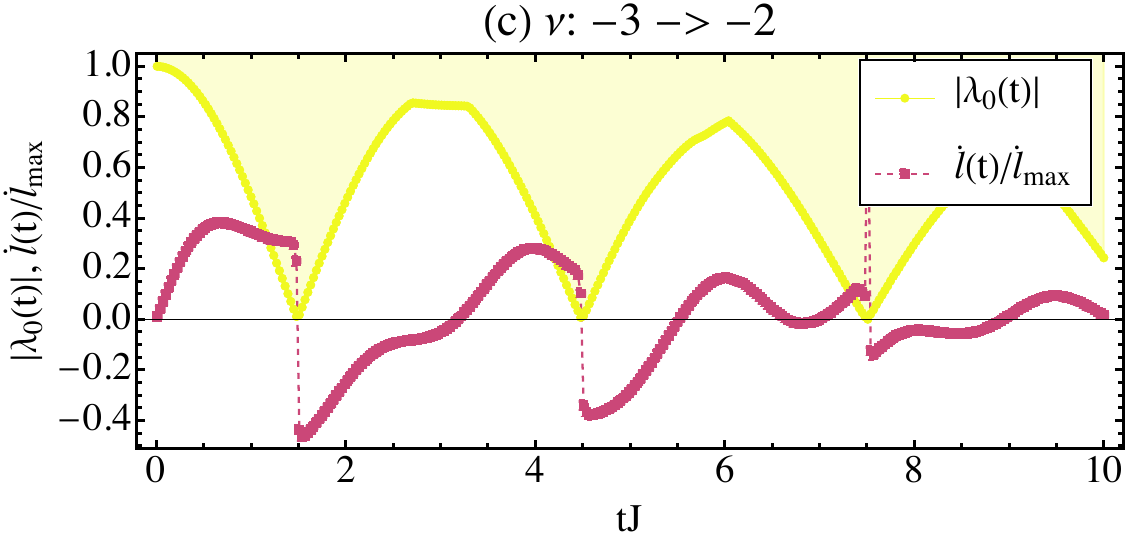}
\includegraphics[width=0.95\columnwidth]{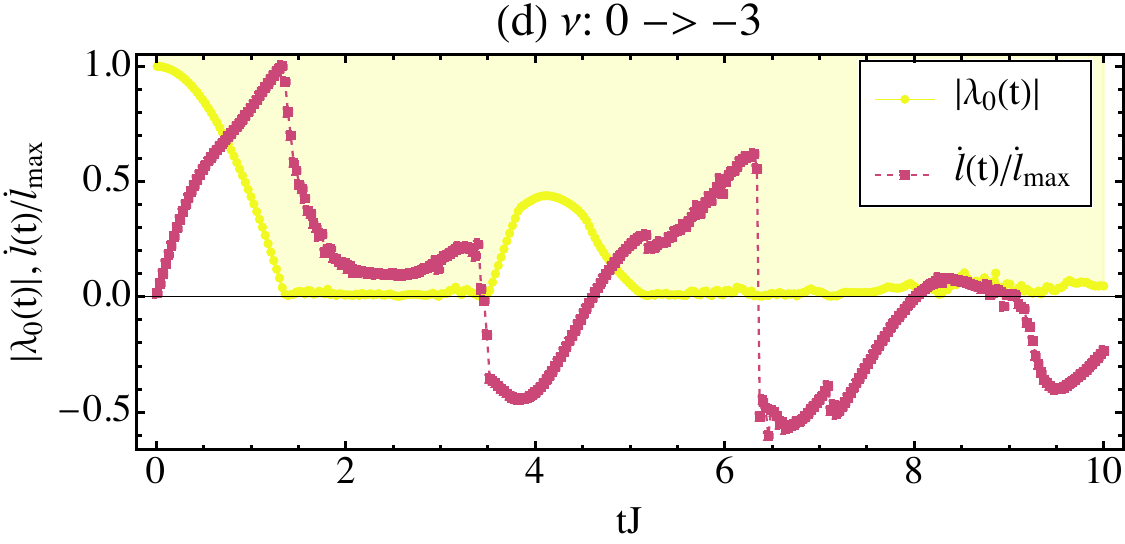}
\caption{Several examples of the derivative of the return rate, $\dot{l}(t)$, for quenches from table \ref{tab:quenches}. The return rate derivative has been rescaled by its maximum value in the plot $\dot{l}_{max}$. The smallest Loschmidt eigenvalues, $|\lambda_0(t)|$, are shown as a proxy for the critical regions which occur when $|\lambda_0(t)|\approx0$. The light yellow shaded regions would be filled by the remaining eigenvalues $\lambda_{n,\vec{k}}(t)$. All results are calculated for a system of size $N=200^2$.}
\label{fig:rrh1}
\end{figure*}

In Fig.~\ref{fig:rrh1} we give examples of the derivative of the return rate, $\dot l(t)$, for some of the quenches in table \ref{tab:quenches}. For three of the chosen examples the critical regions are rather short, with the critical times close together for smaller times $t$. This means that the critical regions do not overlap for the times we look at, making interpretation of the data simpler. In such cases the cusps in $\dot l(t)$, because they are close together, almost look like discontinuities. For the quench $\nu:0\to-3$ the critical regions are longer and the cusps in $\dot l(t)$ are more easily visible, see Fig.~\ref{fig:rrh1}(d). We are not able to directly calculate the critical times for this model and hence we show rather $|\lambda_0(t)|$. When this approaches or leaves zero we infer a critical time, which can be seen to be in agreement with the appearance of cusps in the derivative of the return rate.

\section{Further examples of the dynamical bulk-boundary correspondence}\label{app:dbb}

\begin{figure}
\includegraphics[height=0.45\columnwidth]{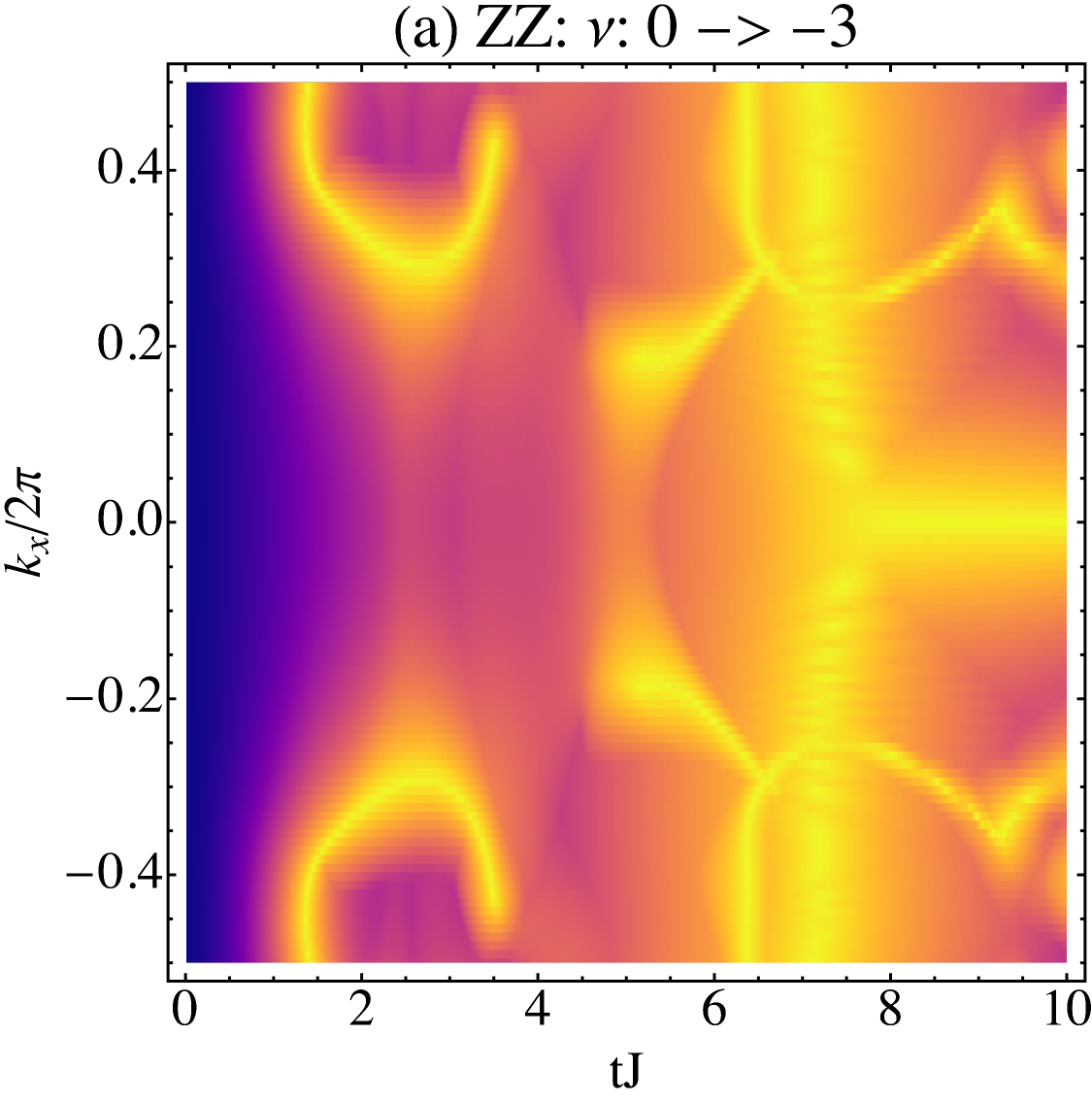}
\includegraphics[height=0.45\columnwidth]{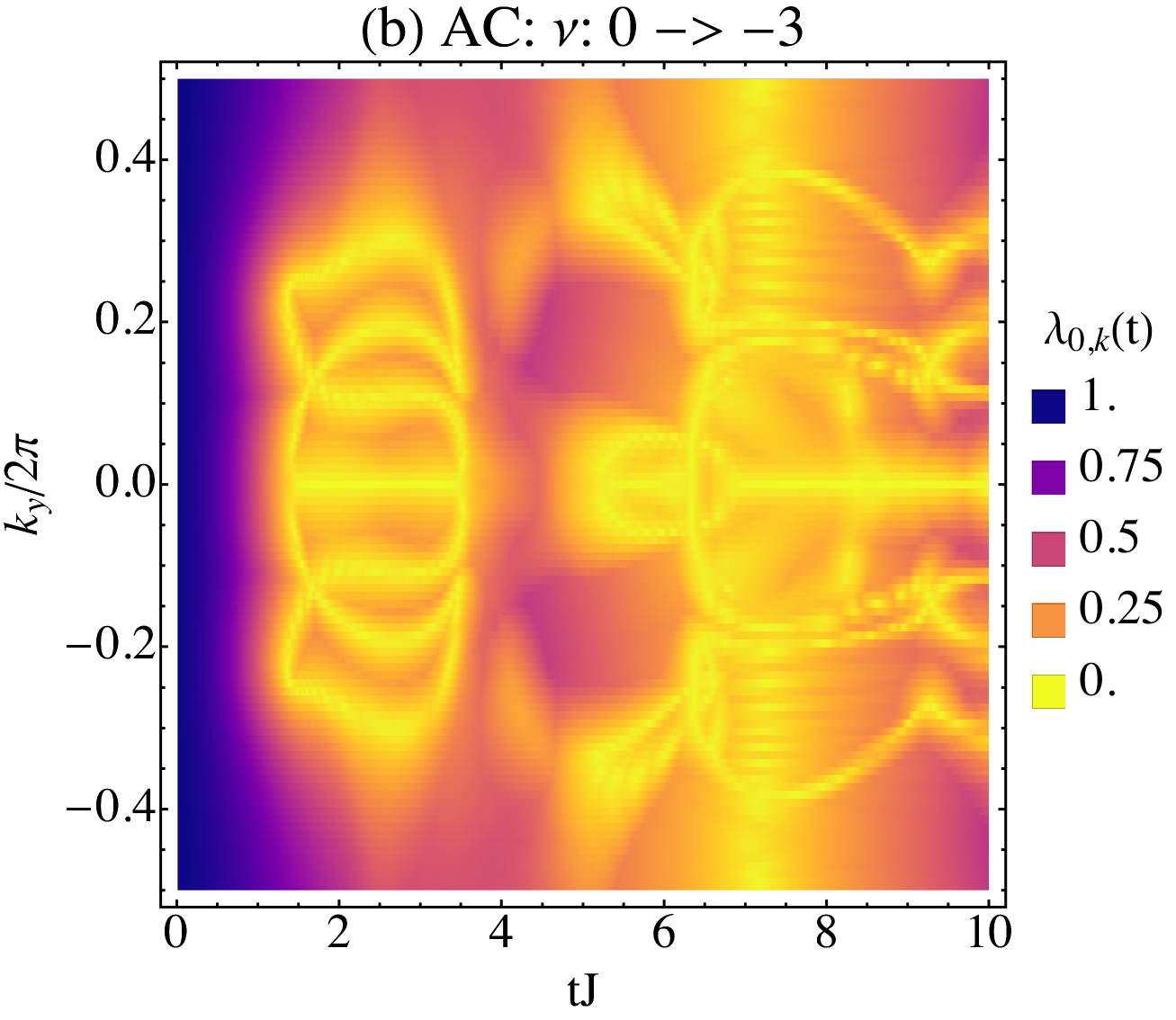}
\caption{In panels (a,b) the lowest Loschmidt eigenvalue $\lambda_{0,k}(t)$ as a function of momentum and time is plotted for both zig-zag (a) and armchair (b) edges. See Fig.~\ref{fig:evrdehex} and table \ref{tab:quenches} for more details.} 
\label{fig:evrdehexapp}
\end{figure}

Finally, in Fig.~\ref{fig:evrdehexapp} we give further examples of $\lambda_{0,k}(t)$ as a function of momentum and time. Here the quench $\nu:0\to-3$ is shown, see table \ref{tab:quenches} for parameters. The patterns of $t_c(\vec{k}^*)$ and the in-gap bands in $\lambda_{0,k}(t)$ giving rise to $|\lambda_{0,k=0}(t)|\approx 0$ are clearly discernible, but again one can see differences for the armchair and zig-zag edges as to when in-gap modes appear in the Loschmidt spectrum.

%

\end{document}